\documentclass[fleqn,twoside]{article}
\usepackage{espcrc2}
\usepackage{latexsym}
\usepackage{euscript}
\usepackage{graphicx}
\usepackage[figuresright]{rotating}

\def\simge{\mathrel{%   
   \rlap{\raise 0.511ex \hbox{$>$}}{\lower 0.511ex \hbox{$\sim$}}}}   
\def\simle{\mathrel{   
   \rlap{\raise 0.511ex \hbox{$<$}}{\lower 0.511ex \hbox{$\sim$}}}}   
\def\slashchar#1{\setbox0=hbox{$#1$}           % set a box for #1   
   \dimen0=\wd0                                 % and get its size   
   \setbox1=\hbox{/} \dimen1=\wd1               % get size of /   
   \ifdim\dimen0>\dimen1                        % #1 is bigger   
      \rlap{\hbox to \dimen0{\hfil/\hfil}}      % so center / in box   
      #1                                        % and print #1   
   \else                                        % / is bigger   
      \rlap{\hbox to \dimen1{\hfil$#1$\hfil}}   % so center #1   
      /                                         % and print /   
   \fi}                                         %   

\def\simge{\mathrel{%   
   \rlap{\raise 0.511ex \hbox{$>$}}{\lower 0.511ex \hbox{$\sim$}}}}   
\def\simle{\mathrel{   
   \rlap{\raise 0.511ex \hbox{$<$}}{\lower 0.511ex \hbox{$\sim$}}}}   
\def\slashchar#1{\setbox0=\hbox{$#1$}           % set a box for #1   
   \dimen0=\wd0                                 % and get its size   
   \setbox1=\hbox{/} \dimen1=\wd1               % get size of /   
   \ifdim\dimen0>\dimen1                        % #1 is bigger   
      \rlap{\hbox to \dimen0{\hfil/\hfil}}      % so center / in box   
      #1                                        % and print #1   
   \else                                        % / is bigger   
      \rlap{\hbox to \dimen1{\hfil$#1$\hfil}}   % so center #1   
      /                                         % and print /   
   \fi}

\newcommand{\AmS}{{\protect\the\textfont2
  A\kern-.1667em\lower.5ex\hbox{M}\kern-.125emS}}

\newcommand{\ba}{\begin{equation} \left\{ \begin{array}{lr}}
\newcommand{\ea}{\end{array} \right. \end{equation}}

\newcommand{\bea}{\begin{eqnarray}}
\newcommand{\eea}{\end{eqnarray}}
\newcommand{\be}{\begin{equation}}
\newcommand{\ee}{\end{equation}}

\newcommand{\fBsmall}{483(4)}
\newcommand{\fB}{170(11)(5)(22)}
\newcommand{\fBstat}{170(11)}

\newcommand{\fBs}{192(9)(5)(24)}

\newcommand{\fBsd}{1.13(2)(1)}
\newcommand{\fDsd}{1.10(1)(1)}
\title{$f_B$ from finite size effects in lattice QCD\thanks{Talk given at Lattice 2002 by R.P.}}

\author{M. Guagnelli\address[ROME2]{Dipartimento di Fisica, Universit\`a di Roma ``Tor Vergata'', 
        V. R. Scientifica 1, I-00133 Rome, Italy}\address[INFN]{INFN Roma 2, V. R. Scientifica 1, I-00133 Rome, Italy}, 
        F. Palombi\addressmark[ROME2]\addressmark[INFN], 
        R. Petronzio\addressmark[ROME2]\addressmark[INFN]
        and
        N. Tantalo\addressmark[ROME2]\addressmark[INFN]}
       
\begin{document}

\begin{abstract}
We discuss a novel method to calculate $f_B$ on the lattice, introduced in \cite{Guagnelli:2002jd}, based on the study of
the dependence of finite size effects upon the heavy quark mass of flavoured mesons and on a non--perturbative 
recursive finite size technique. This method avoids the systematic errors related to extrapolations
from the static limit  or to the tuning of the coefficients of effective Lagrangian and
the results admit an extrapolation to the continuum limit.
We show the results of a first estimate at finite lattice spacing, but close to the continuum limit, giving 
$f_B = \fB$ {\rm MeV}. We also obtain $f_{B_s} = \fBs${\rm MeV}. The first error is statistical, 
the second is our estimate of the systematic error from the method and the third the 
systematic error from the specific approximations adopted in this first exploratory calculation.
The method can be generalized to two--scale problems in 
lattice QCD.
\end{abstract}

\maketitle

%%%%%%%%%%%%%%%%%%%%%%%%%%%%%%%%%%%%%%%%%%%%%%%%%%%%%%%%%%%%%%%%%%%%%%%%%%%%%%%%%%%%%%%%%%%%%%%%%%%%%%%%%%%
\section{Introduction}
%%%%%%%%%%%%%%%%%%%%%%%%%%%%%%%%%%%%%%%%%%%%%%%%%%%%%%%%%%%%%%%%%%%%%%%%%%%%%%%%%%%%%%%%%%%%%%%%%%%%%%%%%%%
Lattice QCD evaluations of quantities characterised by two scales with a large
hierarchy require in general a very high lattice resolution and a sizeable total 
physical volume to correctly account the dynamics of the small distance  scale and to
dispose of the finite size effects related to the large distance  scale. A good example
is provided by the pseudoscalar $B$  meson decay constant~\cite{Ryan:2001ej},  where the small
distance scale is represented by the inverse of the bottom quark mass and
the large distance scale by the radius of the $B$ meson, related in turn to the inverse of the light quark mass.
A straight evaluation of the decay constant would require lattices with $N=80^4$ points or more,
exceeding the present generation computers capabilities, and, in the case
of unquenched simulations, the ones of the next generation.
One resorts to approximate calculations based on extrapolations from the static limit
or on non--relativistic formulations of standard QCD. All the available methods introduce
systematic errors related to extrapolation fits and/or to the use of effective Lagrangians. 

We discuss a novel approach based on the study of the
dependence upon the heavy quark mass of finite size effects for the pseudoscalar decay constant
of heavy flavoured mesons (see \cite{Guagnelli:2002jd} for details). 
The basic assumption is that the finite size effects are mainly related 
to the light quark mass  and rather insensitive to the one of a sufficiently heavy quark.
We discuss the general features of the method assuming the continuum limit has been taken.
The relevant quantity is the ratio $\sigma \equiv f_B( 2L) / f_B(L)$ of the pseudoscalar constants at
different volumes,
where $f_B(L)$ is the value of the decay constant on a volume with linear size $L$.
The dimensionless $\sigma$ depends on general grounds upon three dimensionless
variables: $m_\ell L$, $m_hL$ and $\Lambda_{QCD}L$.  
For a sufficiently large heavy quark mass $m_h$, the dependence is basically dominated by the light quark 
and the expansion for large $m_h$ takes the form
\be
\sigma = \sigma \left( m_\ell L, \Lambda_{QCD}L\right) + \frac{C\left(m_\ell L, \Lambda_{QCD}L\right)}{m_h L} \;.
\label{eq:sigmadef}
\ee
A simple phenomenological ansatz for $\sigma$ can be made based on the concept of
a reduced mass constructed out of the heavy and light quark masses
\be
\sigma = \sigma\left(m_{red} L,\Lambda_{QCD}L\right) \;,
\ee
where $m_{red} = (\mu_1 \mu_2)/( \mu_1 + \mu_2)$.
The quantity $\mu_i$ is a function of the quark mass, but
not only: indeed, for very light masses, finite size
effects are regulated by the physical meson size, which is expected to remain finite when the
light quark mass approches zero. 

A crucial question is the threshold value of the quark mass on a given volume where the
large $m_h$ expansion becomes reliable. As we will show, this value falls in a mass range of the
order of a couple of GeV in the renormalization invariant mass scheme, where the calculation on a
single lattice is affordable.
Under these circumstances, the strategy to obtain $f_B$ is the following. One first
performs a calculation on a lattice where the resolution is suitable for $b$ quark
propagation, but the total volume is unavoidably a small one. This sets $f_B$ on a finite volume.
In order to connect to the large volume results, one needs the step scaling function $\sigma$ 
for values of heavy quark masses generally lower  than those of the simulation where the
finite size value of $f_B$ was obtained.
The possibility of extrapolating $\sigma$ to heavier masses depends upon the validity of the
asymptotic expansion: in a favourable case, as is the real one, one can evaluate the finite
size effects in a reliable way, connecting, by a repeated iteration of the procedure, small volume values of $f_B$
to the ones on large volumes,
\be
f_B^{phys} = f_B(L_0) \sigma(L_0) \sigma(2L_0) \dots \;,
\ee 
and the recursion stops on a volume where $\sigma \simeq 1$ within a required precision.
The continuum limit is obtained by extrapolating to zero lattice spacing the step scaling function
obtained at fixed physical quantities. 
\begin{figure}[h]
\vskip -0.5cm
\includegraphics[width=7cm]{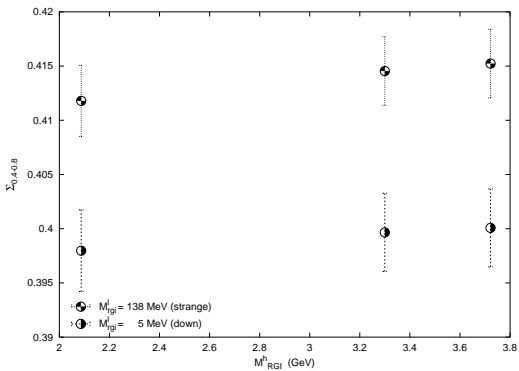}
\vskip -1cm
\caption{\small Step scaling function $\Sigma_{0.4-0.8}$ for the evolution of $f_{h\ell}$ from $0.4$ {\rm fm} to $0.8$ {\rm fm} at $\beta = 6.737$.}
\label{fig:6737}
\vskip -0.4cm
\end{figure}
We use the Schr\"odinger Functional framewrok to compute the pseudoscalar decay constants
and we fix the lattice scale in terms of $r_0$ \cite{Necco:2001xg,toappeartwo}.
Meson masses, on different volumes, are tuned by fixing the physical value of the Renormalization
Group Invariant quak masses \cite{Garden:1999fg,Rolf:2001iv,Heitger:2001ch}.
%%%%%%%%%%%%%%%%%%%%%%%%%%%%%%%%%%%%%%%%%%%%%%%%%%%%%%%%%%%%%%%%%%%%%%%%%%%%%%%%%%%%%%%%%%%%%%%%%%%%%%%%%%%
\section{Results and Discussion}
%%%%%%%%%%%%%%%%%%%%%%%%%%%%%%%%%%%%%%%%%%%%%%%%%%%%%%%%%%%%%%%%%%%%%%%%%%%%%%%%%%%%%%%%%%%%%%%%%%%%%%%%%%%
The results are obtained at finite lattice spacing.
The size of the smallest volume follows from the decision of making our estimate for the  finite size $f_B$ on a $48 \times 24^3$ lattice
with a cutoff of about $a_0^{-1}\simeq 12\ {\rm GeV}$. The value of
the bare coupling for this lattice spacing has been obtained from a
fit in   ref.~\cite{toappeartwo}.
The procedure fixes $\beta(a_0) = 7.3$ and the physical volume $L_0 = 0.4$ {\rm fm}.  
On this lattice, we simulate heavy quark masses up to $0.3$ in lattice units, corresponding
to bare physical masses slightly above $4$~GeV.
Indeed, as a general caution against large lattice artifacts, at all $\beta$ values we take
the maximum heavy quark mass in lattice units of the order of $0.3$.
The first $\Sigma$ (we distinguish between the continuum step function $\sigma$ and
the one at finite lattice spacing $\Sigma$) goes from  the volume of $0.4$ {\rm fm} to the one of $0.8$~{\rm fm}.
In terms of lattice points, we go from $12$ to $24$, and we have to match
the starting volume of $0.4$ {\rm fm} with a resolution which is half of the one used for
a correct estimate of the bottom quark propagation. According to our caveat,
it follows that the maximum  bare  quark mass that we can achieve is correspondingly
halved, i.e. of about a couple of GeV at a bare coupling $\beta = 6.737$. 
We make a further iteration with a second $\Sigma$ going from
$0.8$~{\rm fm} to $1.6$~{\rm fm}, where our investigation of heavy
quark masses stops at the order of the charm quark mass. The corresponding bare coupling is $\beta = 6.211$.
The finite volume effects for this second evolution step are small enough
to make the neglection of the residual volume effects a safe assumption, that however can be tested
explicitly.

The plots in Figs.~1 and~2 show the dependence of $\Sigma$  upon the {\it heavy} RGI quark mass $M_{RGI}^h$ for the two volume jumps and provide
evidence for a plateau of insensitivity to heavy quark masses: the data have been obtained from a linear
extrapolation in $M_{RGI}^\ell$ to the down and strange RGI quark masses reported in \cite{Garden:1999fg}. 
These figures, with the figures of ref.\cite{Guagnelli:2002jd}, support the procedure proposed.
\begin{figure}[hp]
\vskip -0.5cm
\includegraphics[width=7cm]{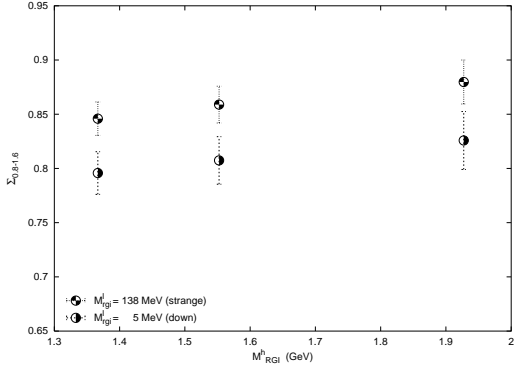}
\vskip -1cm
\caption{\small Step scaling function $\Sigma_{0.8-1.6}$ for the evolution of $f_{h\ell}$ from $0.8$ {\rm fm} to $1.6$ {\rm fm} at $\beta = 6.211$.}
\label{fig:6211}
\vskip -0.4cm
\end{figure}
The statistical errors are computed by a jacknife method (see ref.~\cite{Guagnelli:2002jd} for details).
The finite size value  of $f_B$ is obtained by a calculation on the highest resolution lattice and
the RGI bottom quark mass is obtained from the equation
\be
M = \hat Z_M(g_0) m_{WI}(g_0)
\ee
In order to obtain the renormalisation constant $Z_M(g_0)$  at $\beta = 7.3$ and $\beta = 6.737$, we have used 
a safe interpolation of the pseudoscalar renormalisation constant $Z_P(g_0,\mu)$ at a value of $\mu$ 
three times the reference value used in eq.~(6.8) of ref.~\cite{Capitani:1998mq}. The value for $f_B$ that we obtain is
\be
f_B(0.4{\rm fm} ) = \fBsmall {\rm MeV}
\ee
By using the values of $\Sigma$ for the $b$ quark at constant RGI mass
\be
\Sigma_{0.4-0.8}^{bd} = 0.401(4),\qquad
\Sigma_{0.8-1.6}^{bd} = 0.88(4)
\ee
we obtain our estimate of $f_B$ on the large volume:
\be
f_B^{phys} = \fBstat {\rm MeV} \;,
\ee
where the error quoted in the previous equation is statistical only. 
In this the written version of the talk we include our estimate of the sistematic errors
that can be partly ascribed to specific approximations used in the present computation that 
can be eventually removed, and partly to the uncertainty in the extrapolation in the heavy quark mass of
finite size effects, inherent to the method proposed.
The overall error on the number $f_B$  coming from the removable systematic uncertainties is of about $13$\% and
of at most $2-3$\% from the ones deriving from the unavoidable extrapolation in the heavy quark mass, 
leading to a global uncertainty of about $25$~{\rm MeV} of which about $20$ are removable while $5$ stay with the method:
\be
f_B^{phys} = \fB {\rm MeV} \;.
\ee
In the same way we obtain
\be
f_{B_s}^{phys} = \fBs {\rm MeV} \;.
\ee
We quote also the results of the ratios
\be
\frac{f_{B_s}^{phys}}{f_B^{phys}} = \fBsd \qquad \frac{f_{D_s}^{phys}}{f_D^{phys}} = \fDsd
\ee
that were asked at the end of the talk.

The method proposed can be generalized to problems characterised by two very different mass scales,
if the decoupling of the large mass scale from the low scales of non-perturbative QCD dynamics holds true.
This appears to be the case in the example discussed and is somehow supported by the wide success
of the predictions of perturbative QCD calculations for hard processes that are insensitive to the dressing
mechanism of quarks and gluons into standard hadronic final states.

%%%%%%%%%%%%%%%%%%%%%%%%%%%%%%%%%%%%%%%%%%%%%%%%%%%%%%%%%%%%%%%%%%%%%%%%%%%%%%%%%%%%%%%%%%%%%%%%%%%%%%%%%%%

%%%%%%%%%%%%%%%%%%%%%%%%%%%%%%%%%%%%%%%%%%%%%%%%%%%%%%%%%%%%%%%%%%%%%%%%%%%%%%%%%%%%%%%%%%%%%%%%%%%%%%%%%%%

%%%%%%%%%%%%%%%%%%%%%%%%%%%%%%%%%%%%%%%%%%%%%%%%%%%%%%%%%%%%%%%%%%%%%%%%%%%%%%%%%%%%%%%%%%%%%%%%%%%%%%%%%%%

\begin{thebibliography}{99}
%%%%%%%%%%%%%%%%%%%%%%%%%%%%%%%%%%%%%%%%%%%%%%%%%%%%%%%%%%%%%%%%%%%%%%%%%%%%%%%%%%%%%%%%%%%%%%%%%%%%%%%%%%%

{\small
%\cite{Guagnelli:2002jd}
\bibitem{Guagnelli:2002jd}
M.~Guagnelli, F.~Palombi, R.~Petronzio and N.~Tantalo,
%``f(B) and two scales problems in lattice QCD,''
hep-lat/0206023.
%%CITATION = HEP-LAT 0206023;%%


%\cite{Ryan:2001ej}
\bibitem{Ryan:2001ej}
S.~M.~Ryan,
%``Heavy quark physics from lattice QCD,''
Nucl.\ Phys.\ Proc.\ Suppl.\  {\bf 106} (2002) 86
[hep-lat/0111010].
%%CITATION = HEP-LAT 0111010;%%

%\cite{Necco:2001xg}
\bibitem{Necco:2001xg}
S.~Necco and R.~Sommer,
%``The N(f) = 0 heavy quark potential from short to intermediate  distances,''
Nucl.\ Phys.\ B {\bf 622} (2002) 328
[hep-lat/0108008].
%%CITATION = HEP-LAT 0108008;%%


%\cite{toappeartwo}
\bibitem{toappeartwo}
M.~Guagnelli, R.~Petronzio and N.~Tantalo,
%``The lattice spacing in QCD at large \beta''
hep-lat/0209112.
%%CITATION = HEP-LAT 0209112;%%



%\cite{Garden:1999fg}
\bibitem{Garden:1999fg}
J.~Garden, J.~Heitger, R.~Sommer and H.~Wittig  [ALPHA Collaboration],
%``Precision computation of the strange quark's mass in quenched QCD,''
Nucl.\ Phys.\ B {\bf 571} (2000) 237
[hep-lat/9906013].
%%CITATION = HEP-LAT 9906013;%%


%\cite{Rolf:2001iv}
\bibitem{Rolf:2001iv}
J.~Rolf and S.~Sint,
%``The charm quark's mass in quenched QCD,''
Nucl.\ Phys.\ Proc.\ Suppl.\  {\bf 106} (2002) 239
[hep-ph/0110139].
%%CITATION = HEP-PH 0110139;%%

%\cite{Heitger:2001ch}
\bibitem{Heitger:2001ch}
J.~Heitger and R.~Sommer  [ALPHA collaboration],
%``A strategy to compute the b quark mass with non-perturbative accuracy,''
Nucl.\ Phys.\ Proc.\ Suppl.\  {\bf 106} (2002) 358
[hep-lat/0110016].
%%CITATION = HEP-LAT 0110016;%%

%\cite{Capitani:1998mq}
\bibitem{Capitani:1998mq}
S.~Capitani, M.~Luscher, R.~Sommer and H.~Wittig  [ALPHA Collaboration],
%``Non-perturbative quark mass renormalization in quenched lattice QCD,''
Nucl.\ Phys.\ B {\bf 544} (1999) 669
[hep-lat/9810063].
%%CITATION = HEP-LAT 9810063;%%

}
\end{thebibliography}
\end{document}